\documentstyle[aaspp4]{article} 

\newcommand\simlt{\lower.5ex\hbox{$\; \buildrel < \over \sim \;$}}
\newcommand\simgt{\lower.5ex\hbox{$\; \buildrel > \over \sim \;$}}

\begin{document}
\title{On time evolution and causality of force-free black hole magnetospheres}
\author{Amir Levinson\altaffilmark{1,2}}
\altaffiltext{1}{School of Physics \& Astronomy, Tel Aviv
University, Tel Aviv 69978, Israel; Levinson@wise.tau.ac.il}
\altaffiltext{2}{School of Physics, University of Sydney, NSW 2006}
\begin{abstract}
The rotational energy of a Kerr black hole is often invoked as the free energy source 
that powers compact astrophysical systems.
An important question concerning the energy extraction mechanism is
whether a Kerr black hole can communicate stresses to a distant load via a 
surrounding, force-free magnetosphere, and whether such structures are stable. 
In this paper we address this question.
We first re-examine the properties of short wavelength force-free waves,
and show that contrary to earlier claims, fast magnetosonic disturbances do affect the 
global electric current flowing in the system, even in flat spacetime; beyond the light
cylinder the perturbed charge density is on the order of the perturbed 
Goldreich-Julian charge density.  This implies 
that the fast magnetosonic surface is a causal wind boundary.
We then go on to study the evolution 
of a force-free system driven by a spinning-up black hole, by solving 
the time dependent Maxwell equations near the horizon using 
perturbation method.  We find that the electromagnetic field,
current and charge density grow exponentially in the linear regime.  We conclude that after 
time $t=2M\ln(1/2\alpha^2)$, where $\alpha$ is the lapse function, the evolution will reach an
adiabatic stage, whereby the solution is to a good approximation 
the static one, with the angular momentum of the hole being the adiabatic parameter.
This implies that black hole
magnetospheres are stable.  A brief discussion on the relation
between global current closure and causality in force-free systems is given at the end.
\end{abstract}


\section{Introduction}

Energy extraction from a Kerr black hole by a magnetized inflow 
is believed to be a plausible mechanism for powering compact astrophysical
systems. The magnetic field lines threading the horizon may be anchored
to a disk or a torus surrounding the black hole, where the 
extracted energy dissipates, or may extend to infinity, as envisioned in some
applications of this mechanism to jet formation in AGNs, GRBs, and microquasars,
in which case dissipation occurs at some distant load.

Energy extraction along a magnetized flux tube that thread the horizon occurs when 
(i) the angular velocity associated with the flux tube is larger than zero 
and smaller than the hole angular velocity, and (ii) the Alfven point of the 
inflow is located inside the ergosphere (Takahashi et al. 1990). 
In the original model proposed by Blandford \& Znjek (1977; hereafter BZ77), as
well as in some other versions (e.g., van Putten 2001),
it is conjectured that the hole rotational energy (or a fraction of it, as in van 
Putten's model) is extracted along an open, force-free flux 
tube that extends from the horizon to well beyond the outer light cylinder.  In the 
context of ideal MHD it is clear that there must exist a region between the inner
and outer light cylinders where ideal MHD is violated, since the streamlines change 
direction (matter is inflowing into the horizon and outflowing to infinity).  This region
serves as a plasma source, e.g, via pair creation in a sparking gap.  In the force free
limit, in which inertia is neglected, this deviation is thought to be sufficiently small,
to allow the force free approximation to be valid in the entire region between 
the horizon and the load - the region above the outer light cylinder, 
where the extracted energy dissipates (e.g., Blandford 2002).  

The question whether a force-free magnetosphere can exist and whether it is stable
has been the subject of a recent debate (Punsly \& Coroniti 1990, hereafter PC90; 
Beskin \& Kuznetsova 2000;  Komissarov 2001; Blandford 2001, 2002; Punsly 2003, van Putten 
\& Levinson 2003).  In particular it has been argued that a force free black hole 
magnetosphere is not a causal structure and, therefore, physically excluded.
The point is that  in the force free limit the electric current cannot flow across poloidal 
magnetic field lines, and is therefore conserved on magnetic flux surfaces. As a consequence
the angular velocity, $\Omega_F$, of a flux tube extending from the horizon is not a free 
parameter, but is determined by matching boundary conditions on the horizon and at infinity
(see BZ77; Phinney 1983).  
This, according to PC90, violates the 
principle of MHD causality, because the inflowing magnetic wind must pass through the inner 
light cylinder before reaching the horizon and, therefore, cannot communicate 
with the plasma source region [e.g., the gap in the Blandford-Znajek model, or the 
torus in the model proposed by van Putten (2001; see also van Putten \& Levinson 2003)].
PC90 concluded that the use of the Znajek frozen-in condition on the horizon
to determine $\Omega_F$ is unphysical, and that $\Omega_F$ must be determined 
by the dissipative process that leads to ejection of plasma on magnetic field 
lines between the inner and outer Alfven points. 

Blandford (2001, 2002) claimed that a fast magnetosonic mode can propagate across magnetic
field lines at the speed of light and carry information about the toroidal magnetic 
field and poloidal current to the plasma source even beyond the light cylinder.  In 
the force free limit the fast critical surface approaches the horizon, and so stresses
can be communicated to distant regions.  Beskin \& Kuznetsova (2000) reached similar conclusions by 
analyzing the Grad-Shafranov equation.
A recent analysis of force free waves has been carried out by Punsly (2003),
who has shown that linear perturbations of field aligned current and charge density 
cannot propagate along the fast wave characteristics.  He therefore concluded that the 
use of the event horizon as a boundary surface in the force free limit is physically 
not allowed.  The analysis of Punsly assumes the existence of a frame where the electric 
field vanishes and where Maxwell equations do not change form.  We question this later assumption.
In a frame rotating with the flux tube, for example, the electric field indeed 
vanishes.   However, this frame  
is noninertial and Maxwell's equations should be properly transformed.  In particular, 
the unperturbed charge density does not vanish in this frame and should be accounted for 
in the perturbed force-free condition for the waves. We discuss this point further in appendix A.1.
Moreover, in Kerr spacetime proper account must be taken for the effect of 
frame dragging.   Detailed investigation of force-free waves in Kerr spacetime is presented in 
Uchida (1997).  Unfortunately, it does not address these issues in a clear way.
Below, we re-analyze short wavelength disturbances and
show that charge and current perturbations are in fact induced by fast magnetosonic disturbances,
even in flat spacetime, and that beyond the light cylinder the perturbed charge density is of the
order of the perturbed Goldreich-Julian (GJ) charge density.
We also argue that frame dragging plays essentially a similar role as a unipolar inductor.  This
has already been pointed out earlier by  Beskin \& Kuznetsova (2000).
To elucidate the effect of frame dragging on the time evolution of a black hole magnetosphere,
we examine, in the second part of this paper, how a force free
system evolves in the presence of a spinning-up black hole, by solving 
the time dependent Maxwell equations near the horizon using 
perturbation method.  We find an exponential growth of toroidal magnetic field, current 
and charge density in the linear regime, and conclude that a black hole magnetosphere 
is stable.  Our results appear to be consistent with recent numerical simulations 
(Komissarov 2001; the rapid growth of the toroidal magnetic field found in the MHD 
simulations performed by Koide et al. [2002], and Koide [2003] is also consistent 
with the results described in sec. 2), and can also serve as a test case for such 
simulations.

\section{Analysis of force-free disturbances}
In the 3+1 formalism (Thorne, Price and Macdonald 1986; Beskin 1997)
Maxwell's equations can be written, in terms of the electric and magnetic fields,
and charge and current densities measured by a ZAMO, as
\begin{eqnarray}
\frac{1}{c}\frac{\partial {\bf B}}{d t}+{\bf \nabla}\times(\alpha{\bf E})=-({\bf B}\cdot
{\bf \nabla}\beta){\bf m},\label{M3a}\\
-\frac{1}{c}\frac{\partial {\bf E}}{d t}+{\bf \nabla}\times(\alpha{\bf B})=\frac{1}{c}
4\pi \alpha {\bf j}+({\bf E}\cdot{\bf \nabla}\beta){\bf m},\label{M3b}\\
{\bf \nabla}\cdot{\bf B}=0,\label{M3c}\\
{\bf \nabla}\cdot{\bf E}=4\pi\rho_e,\label{M3d}
\end{eqnarray}
where the lapse function $\alpha$, and the ZAMO angular velocity  $-\beta$, are given explicitly in appendix B.
Here ${\bf m}=g_{\phi\phi}{\bf \nabla}\phi$ denotes the associated Killing vector field.  In terms of 
the Boyer-Linduist coordinates it is given as ${\bf m}=\Sigma\sin\theta/\rho\hat{\phi}$, where the metric terms $\rho$
and $\Sigma$ are defined in appendix B.  We denote the components of a vector
in our local coordinate system by ${\bf A}=(A_1,A_2,A_3)$, and choose $A_3$ to be the toroidal component 
(that is, in the direction of ${\bf m}$).  
The force free condition takes the same form as in flat spacetime, viz.,
\begin{equation}
\rho_e{\bf E}+{\bf j}\times{\bf B}/c=0;\ \ \ \ {\bf E}\cdot{\bf B}=0.
\label{FF3+1}
\end{equation}
From eq. (\ref{M3a}) it is clear that since the ZAMO angular velocity is not conserved on magnetic flux 
surfaces, it introduces a coupling between the poloidal and toroidal components of the magnetic 
field.  Furthermore, this term cannot be transformed away.  The effect of this coupling on the 
evolution of the system is particularly important when the spacetime itself is changing with time.  
For instance, consider a situation wherein the angular momentum of the black hole 
undergoes low amplitude time variations,
viz., $a(t)=a_0+\delta a(t)$, with $\delta a<<a$.  To first order in $\delta a$, terms of the form 
$-{\bf m}({\bf B}\cdot{\bf \nabla})\delta {\beta}(t)$, where ${\bf B}$ is the unperturbed magnetic field, 
and likewise for ${\bf E}$, would appear in the wave equation for the perturbed electromagnetic field.
These terms act as a driving force that excites waves, much like a Faraday disk with perturbed 
angular velocity, with one difference, that the driving force associated with frame dragging acts 
everywhere in space, not only on the horizon surface, although it declines steeply with radius.
This qualitative discussion illustrates that spacetime does act like a unipolar inductor through 
the long range gravitomagnetic force, and that changes in spacetime are partly communicated 
to the plasma source by gravity.   In the next section we present a quantitative treatment of 
this problem.  

The question whether the fast critical surface is a causal wind boundary still remains nonetheless,
and is of relevancy also to super Alfvenic outflows in flat spacetime, as in the pulsar case.
As we now show, if the unperturbed flux tube is rotating, then charge perturbations do
propagate along cross-field fast characteristics even in flat spacetime.  To elucidate this
point we shall, in what follows,
consider linear perturbations of a static solution of equations (\ref{M3a})-(\ref{FF3+1})
far from the black hole.  We can then set $\alpha=1$ and $\beta=0$.  
Now, the unperturbed charge density approaches asymptotically the GJ value, that is,
$\rho_e\simeq \rho_{GJ}=E_2\cos\theta/2\pi R$, where $R$ is a cylindrical radius measured with 
respect to the rotation axis and $\theta$ is the angle between the rotation axis and $\hat{e}_1$, 
and so in the force-free condition ({\ref{FF3+1})
the terms ${\bf j}\times\delta {\bf B}$ and $\rho\delta{\bf E}$ are smaller by a factor of $(kR)^{-1}$
than the terms  $\delta {\bf j}\times{\bf B}$ and $\delta\rho_e{\bf E}$.  This means that to 
zeroth order these terms can be neglected.  As we shall show, the zeroth order solution 
gives the dispersion relations and eigen modes.  However, it is clear that in order to maintain
the asymptotic charge density near the GJ value, perturbations 
on the order of $\delta\rho_e\sim\delta E_2/R$ are required; that is, the perturbed charge and 
current density appear only to first order.  
We shall therefore keep the terms associated with the unperturbed charge and current density 
in our analysis.  We consider short-wavelength disturbances, and employ the WKB 
approximation, whereby the rapidly oscillating part of the perturbed 
quantities assumes the form $\exp i[\psi({\bf r})-\omega t]$.  We define the wave vector of the 
disturbances as usual to be ${\bf k}={\bf \nabla}\psi$.
We then obtain for the perturbed quantities:
\begin{eqnarray}
(k^2c^2-\omega^2)\delta{\bf E}-c^2{\bf k}({\bf k}\cdot\delta {\bf E})=i4\pi\omega\delta{\bf j},\label{pert1}\\
c{\bf k}\times \delta{\bf E}=\omega\delta{\bf B},\label{pert2}\\
4\pi \delta \rho_e=i{\bf k}\cdot \delta {\bf E},\label{pert3}\\
c{\bf E}\delta\rho_e+c\rho_e\delta {\bf E}+\delta{\bf j}\times{\bf B}+{\bf j}\times \delta{\bf B}=0,\label{pert4}\\
{\bf B}\cdot\delta{\bf E}+{\bf E}\cdot\delta{\bf B}=0. \label{pert5}
\label{Disp}
\end{eqnarray}
These equations can be cast into the form $\Lambda_{ij}({\bf k},\omega)\delta E_j=0$.  The matrix 
$\Lambda_{ij}({\bf k},\omega)$ is derived for axisymmetric modes, viz., $k_3=0$, in appendix A.
In order to have a nontrivial solution, the determinant of $\Lambda_{ij}$ must vanish.  The latter
is given by eq. (\ref{det}).  We find no corrections to first order to the 
dispersion relations of the intermediate (see eq [\ref{intermediate})]), and fast ($\omega=kc$)
modes.  In the remaining part of this section we analyze only the fast mode.  The perturbed 
charge and current density of the fast mode vanish to zeroth order 
as inferred by Punsly (2003).  However, using eq. (\ref{rho-wav}) we find that to first order 
the charge density is given by,
\begin{equation}
\delta \rho_e=-\frac{k_{||}j_p-k\rho_ec}{k_1c[B_3+(\Omega_F/\omega)Rk_{||}B_p]}\delta E_2,
\label{rho-wav2}
\end{equation}
where $B_p$ denotes the poloidal magnetic field, $j_p$ the poloidal current, 
and $k_{||}$ the component of the wave vector along $B_p$.
It is instructive to examine its behavior in the super-Alfvenic regime.
Beyond the outer light cylinder the unperturbed poloidal current becomes purely convective, viz.,
${\bf j}_p=\rho_e{\bf v}_p$, and the toroidal magnetic field approaches $B_3=E_2$ (e.g, Mestel 1999).
Near the axis we therefore have $B_3\simeq-\Omega_FRB_p/c$ and $k_{||}\simeq k_1$.  Substituting
these results into eq. (\ref{rho-wav2}) yields,
\begin{equation}
\delta \rho_e\simeq \frac{k\rho_e}{k_{||} E_2}\delta E_2\simeq \frac{\delta E_2}{2\pi R}\simeq\delta \rho_{GJ}.
\end{equation}
Thus, the perturbed charge density is approximately the perturbed GJ density beyond the 
light cylinder.  This result follows 
essentially from the requirement on the wave to be force-free.  Since 
$\omega\delta \rho_e=k_{||}\delta j_p$, it follows that {\it perturbations near the fast critical surface
will affect the global current flowing in the system and will be communicated to the source}.
The solution for the perturbed magnetic field can be found now from eq. (\ref{dE2}) and (\ref{pert2}).
In the region above the light cylinder it approaches
\begin{equation}
\delta E_2\simeq -\frac{k_{||}}{k}(1+k_{||}/k)(\Omega_FR/c)\delta B_1.
\end{equation}
The analysis near the horizon is somewhat more involved, and will not be presented here. 
Recall, however, that the force-free condition (\ref{pert4}) is valid also near the horizon
in the ZAMO frame.  By taking the projection of eq. (\ref{pert4}) on the $\hat{e}_3$
direction we obtain:
\begin{equation}
(\rho_e-{\bf k}\cdot{\bf j})\delta E_3+\delta {\bf j}\times{\bf B}=0.
\end{equation}
Since $\delta E_3\ne 0$ it is clear that current perturbations will be induced also 
in the vicinity of the horizon.  However, in order to calculate the perturbed charge density,
a solution for the eigen vector of the fast mode near the horizon must be found first.
Our conclusion is that in cases in which the force-free limit is approached, the angular velocity
of a rotating flux tube threading the event horizon of a Kerr black hole 
will be determined by the fast critical surfaces.  

\section{Spacetime driven evolution of a force free magnetosphere}

Consider a black hole, initially non-rotating, embedded in an 
asymptotically uniform magnetic field directed along the rotation axis of the hole.  We 
suppose that initially the magnetic field is described by the vacuum solution of 
eqs. (\ref{Max t})-(\ref{Max theta}), which is given by 
$A_t=F_{r\theta}=0$, and 
\begin{equation}
A_\phi=(B/2)r^2\sin^2\theta.
\label{Aphi-0}
\end{equation}
We further assume that the space surrounding the hole is filled with a tenuous plasma 
that maintains the system in a force-free state.
Imagine now that the state of the black hole is perturbed adiabatically, such that 
it slowly acquires angular momentum, and that to a good approximation the spacetime can 
still be described by the Kerr metric with $a=a(t)$.  These temporal variations of spacetime
will induce perturbations in the electromagnetic field.  For simplicity we consider here the
regime of slow rotation, viz., $a/M<<1$.  For convenience we use the fields $\psi_\theta$,
$\psi_r$, $B_T$, and $A_\phi$ defined in appendix B as our independent variables. 
To derive the equations for the perturbed electromagnetic field, we linearize eqs 
(\ref{M1})-(\ref{FF4}), using the hole
angular momentum $a$ as the small parameter.  To zeroth order $\psi_\theta=\psi_r=B_T=0$, and 
$A_\phi$ is given by eq. (\ref{Aphi-0}).  The components of the metric tensor satisfy
$\alpha^2=1-2M/r +0(a^2/M^2)$;  $\Delta=r^2-2Mr+a^2$, $\rho^2=r^2+a^2\cos\theta^2$; 
and $\beta =-2aM/r^3+0(a^3/M^3)$.  From eqs (\ref{FF3}) it is clear the $\tilde{j}^\phi$
is of order $0(a^2/M^2)$, and that to second order in $a/M$ 
the force-free conditions (\ref{FF1}) and (\ref{FF2}) reduce to,
\begin{eqnarray}
j^r=-\frac{A_{\phi, \theta}}{A_{\phi, r}}j^\theta=-r\cot\theta j^\theta,\label{FFb}\\
\Delta\psi_{\theta}=\frac{A_{\phi,\theta}}{A_{\phi,r}}\psi_r=r\cot\theta\psi_r.
\label{FFa}
\end{eqnarray}
From equations (\ref{FFb}), (\ref{FFa}), (\ref{M1}), (\ref{M2}) and (\ref{M5}), we obtain a differential equation 
for the field $\psi_\theta$:
\begin{equation}
(1+\Delta\tan^2\theta/r^2)\psi_{\theta,t,t}-\frac{1}{r^2\sin\theta}(B_{T,t})_{,r}+
\frac{1}{r^3\cos\theta}(B_{T,t})_{,\theta}=0,
\label{eq-psi}
\end{equation}
with
\begin{equation}
B_{T,t}=\frac{\Delta\sin\theta}{r^2}\left[(\Delta\psi_\theta)_{,r}
-\frac{\Delta}{r}(\tan\theta\psi_{\theta})_{,\theta}
-\frac{6MBa}{r^2}\sin\theta\cos\theta\right].
\label{B_Tt}
\end{equation}
From eq. (\ref{B_Tt}) it is evident that the system is driven initially by the change of
spacetime, specifically by the differential frame dragging term $\beta_{,r}F_{\phi\theta}
=-(6MBa/r^2)\sin\theta\cos\theta$.  No evolution would ensue (to first order in $a$) 
in the absence of this term.  In fact, it is easy to show that the coupling of the 
electromagnetic field to spacetime to first order in $a$ is a consequence of the fact 
that the ZAMO rotation velocity $-\beta$ is not conserved on magnetic flux surfaces.
The above set of equations can be solved to yield the perturbed poloidal electric field
and current, and toroidal magnetic feld, to second order in $a/M$.  The charge density 
is then given by,
\begin{equation}
j^{t}_{,t}+\frac{1}{r^2}(r^2 j^r)_{,r}
+\frac{1}{\sin\theta}(\sin\theta j^\theta)_{,\theta}=0.
\end{equation}
Finally, linearizing eq (\ref{M3}), we derive an inhomogeneous wave equation for the 
toroidal component of the perturbed vector potential $\delta A_\phi$:
\begin{equation}
\delta A_{\phi,tt} +\frac{\Delta}{r^2}
\left[\frac{\Delta}{r^2}\delta A_{\phi,r}\right]_{,r}+\sin\theta
\left[\frac{\Delta}{r^4\sin\theta}\delta A_{\phi,\theta}\right]_{,\theta}=4\pi\Delta \tilde{j}^\phi,
\label{Aphi}
\end{equation}
with $\tilde{j}^\phi$ given by eq. (\ref{FF3}).  
At $t=0$ all perturbed quantities are zero, which defines the initial conditions.

In the following we take for illustration $a(\tau)=a_0[1-\exp(-\tau/T)]$, where we define the 
dimensionless time $\tau=t/2M$.
Near the horizon, the dimensionless variable $x=\alpha^2/2=\Delta/2M^2$ is small.
Eq. (\ref{eq-psi}) can be solved using perturbation approach.
We suppose that the solution can be expanded as 
\begin{equation}
\psi_\theta=\Sigma_n{f_n(\tau,x,\theta)},
\end{equation}
where $f_n$ is of order $x^n$.  To zeroth order we obtain
\begin{eqnarray}
\frac{\partial^2 f_0}{\partial\tau ^2}-f_0=\frac{3Ba}{4M^2}\sin\theta \cos\theta,\\
B_{T,\tau}=x2M^2\sin\theta\left(f_0-\frac{3Ba}{4M^2}\right).
\end{eqnarray}
The solution that satisfies the initial conditions $f_0=B_T=0$ at $\tau=0$ is
\begin{equation}
f_0(\tau,x,\theta)=\frac{3Ba_0\sin2\theta}{8M^2(T^2-1)}\left[-T^2
e^{-(\tau/T)}-T\sinh\tau+\cosh\tau+T^2-1\right]. 
\end{equation}
There is no contribution to the electric currents $j^r$ and $j^\theta$ to this order.
To the next order we find 
\begin{eqnarray}
f_1(\tau,x,\theta)=f_0 x-\frac{3Ba_0\sin2\theta }{8M^2(4T^2-1)}\left[2T^2(1-
e^{-\tau/T})+\frac{1}{2}(\cosh2\tau-1)-T\sinh2\tau\right]x,\\
B_{T1}=2M^2\sin\theta x\int_0^\tau{(2f_1-f_0 x)d\tau^\prime},\\
j^{r}_1=-\frac{3Ba_0\cos^2\theta}{4M^2}x\left
[\frac{T}{T^2-1}(\cosh\tau-e^{-\tau/T})-\frac{\sinh\tau}{T^2-1}+\frac{6T}{4T^2-1}(1-e^{-\tau/T})\right].
\end{eqnarray}
As seen, the perturbed force free field grows exponentially with an e-folding time $2M$.
It is clear that after time $t\sim 2M\ln(2M^2/\Delta)$ the term $f_1$ becomes comparable
to $f_0$, and the above solution is no longer valid.  We conjecture
that after this time the evolution becomes adiabatic, in the sense that at any given time, $\tau$, 
the solution is given to a good approximation by the steady state solution with $a=a(\tau)$ being
the adiabatic parameter.
Note that after time $t\sim 2M\ln(2M^2/\Delta)$ the toroidal magnetic field and charge density
evolve close to their steady-state values, viz., $B_T\sim (\Omega_H/2)F_{\phi\theta}\sin\theta$, and  
$\rho_e=\alpha^2 j^t\sim \Omega_H B_r\cos\theta$, where $\Omega_H=a/4M^2 +0(a^3/M^3)$ 
is the angular velocity of the black hole and $B_r$ is the radial magnetic field.  
A more complete treatment will be presented elsewhere.  The above suggests that the black 
hole magnetosphere is stable; the force-free plasma adjusts quickly to changes in spacetime.

\section{Discussion}
Based on the results found in sections 2 and 3, we argue that a force-free, black hole magnetosphere
is globally causal, and that frame dragging plays an essential role in establishing such 
structures.  We argue that the force-free limit is a good approximation to ideal MHD systems
when plasma inertia becomes negligible.  Our analysis elucidates how the angular 
velocity of magnetic flux tubes penetrating the horizon responds to changes in spacetime, and
confirms recent results found in numerical simulations (Komissarov 2001; Koide 2003).  Moreover, 
the results of section 3 imply that a force-free magnetosphere is stable.
Similar conclusions have been drawn by Beskin \& Kuznetsova (2000), who present an elegant 
analysis of critical surfaces using the Grad-Shafranov equation.  They show that the 
frozen-in condition on the horizon is an integral of the Grad-Shafranov equation, and 
contains no additional information.  van Putten \& Levinson confirm this point by showing that
the frozen-in condition for a cold MHD inflow is merely a consequence of the fact that 
all trajectories approache those of a free-falling observer on the horizon.    
Whether the conditions required for
a stationary force-free plasma to exist near a black hole are satisfied in nature
is a different problem, that may involve issues concerning the microphysics of the plasma source, 
cross field diffusion, entrainment
of matter by instabilities, etc.  The other critical issue, yet to be resolved, concerns the 
global current closure in such systems.  A particular example of a current circuit is discussed
in van Putten \& Levinson (2003). 

Even in flat spacetime one should be cautious in analyzing global
causality.  The analysis presented in Sec. 2 demonstrates that changes in the 
state of the system beyond the light cylinder (but not beyond the fast critical 
surface) can directly affect the global current flowing in the system.  Moreover,
current closure is a critical issue in such systems, and until it is properly 
modeled the question how the system adjusts to changes will remain unresolved.  The wind 
region may be but one section of the global system; the return current may flow in subcritical 
regions, e.g., a cocoon surrounding a fast jet, and these regions may also affect the response 
of the system, regardless of our conclusion above.  Recent analysis (Goodwin et al 
2003) indeed suggests that boundary conditions beyond the light cylinder may considerably 
influence the properties of the wind in a pulsar.  
The response of the system may involve some feedback on the gap in relevant cases.
In my view, sparking gaps as those envisioned in pulsar and black hole models are likely to have
oscillatory behavior because steady state requires fine tuning of the microphysics.  This 
may have important consequences for the properties of these systems that are yet to be explored. 

I thank J. Bekenstein, V. Beskin, Q., Luo,  Y. Lyubarsky, D. Melrose and 
M. van Putten for useful discussions.  This research was supported by 
an ISF grant for a Israli Center for High Energy Astrophysics.

\appendix
\section{Solution of the wave equations}

Equations (\ref{pert2})-(\ref{pert4}) can be solved for the perturbed current and magnetic field.  We 
denote by $B_p$ and $j_p$ the unperturbed poloidal magnetic field and current density 
($j_p^2=j_1^2+j_2^2$), and by $\theta_j=j_p/B_p$ their ratio.  Using
the relations between the unperturbed quantities: $E_1=(\Omega_FR/c)B_2$, and $E_2=-(\Omega_FR/c)B_1$, and 
$j_3=j_pB_3/B_p-\rho_eE_2/B_1$, where $\Omega_F$ is the 
angular velocity of the flux tube and $R$ is a cylindrical radius, and the fact that the 
poloidal current flows along poloidal magnetic field lines, we can express the result 
in terms of the unperturbed magnetic field, charge density and poloidal current 
as: $\delta B_i= A_{ij}\delta E_j$, and  $\delta j_i= D_{ij}\delta E_j$, where,

\begin{eqnarray}
A_{1j}=(k_2c/\omega)\delta_{3j},\\
A_{2j}=-(k_1/k_2)A_{1j},\\
A_{3j}=-(k_2c/\omega)\delta_{1j}+(k_1c/\omega)\delta_{2j}\\
D_{1j}=-\frac{B_1\omega^2}{k_1B_1+k_2B_2}k_j+i4\pi\theta_j\omega A_{1j}-
\frac{i4\pi\omega\rho_e k_2}{k_1B_1+k_2B_2}\delta_{3j},\\
D_{2j}=-(k_1/k_2)D_{1j}-\frac{\omega^2}{k_2}(k_1\delta_{1j}+k_2\delta_{2j}),\\
D_{3j}=(B_3/B_1)(D_{1j}-i4\pi\omega\theta_j A_{1j})+i4\pi\omega(\theta_jA_{3j}-\delta_{2j}\rho_e/B_1)
-(\omega\Omega_FR)(k_1\delta_{1j}+k_2\delta_{2j}).
\end{eqnarray}
Substituting the above results into eq (\ref{pert1}), and using eq. (\ref{pert5}), yields 
a set of equations for the perturbed electric field: $\Lambda_{ij}\delta E_j=0$, where 
the matrix is given by 
\begin{eqnarray}
\Lambda_{1j}=B_j+(\Omega_F R/\omega)(k_1B_1+k_2B_2)\delta_{3j},\\
\Lambda_{2j}=(k^2c^2-\omega^2)\delta_{2j}-k_2k_jc^2-D_{2j},\\
\Lambda_{3j}=-D_{3j}+(k^2c^2-\omega^2)\delta_{3j}.
\end{eqnarray}
Let us denote the poloidal magnetic field by $B_p$, and the components of the wave vector parallel
and perpendicular to the poloidal magnetic field by $k_{||}$ and $k_{\perp}$, respectively.  
We can then write, $k_1B_1+k_2B_2=k_{||}B_p$, and $k_1B_2-k_2B_1=-k_{\perp}B_p$.
The determinant of $\Lambda_{ij}$ then takes the form, 
\begin{eqnarray}
\Delta=\frac{k_1(\omega^2-k^2c^2)}{k_{||}B_p}
\left\{(k_{||}^2c^2-\omega^2)B_p^2+(B_3\omega+\Omega_FRk_{||}B_p)^2\right\}.\label{det}
\end{eqnarray}
The requirement $\Delta=0$ then yields the dispersion relations.  
We recover the familiar result; there are two modes, the intermediate mode satisfying,
\begin{equation}
\omega=\frac{k_{||}cB_p}{B_3^2-B_p^2}\left\{-B_3(\Omega_FR/c)\pm
\left[B_p^2(1+\Omega_F^2R^2/c^2)-B_3^2\right]^{1/2}\right\},
\label{intermediate}
\end{equation}
and having its group velocity directed along poloidal field lines, and the fast 
mode which satisfies $\omega=kc$.  There are no higher 
order corrections to the dispersion relations.  We now proceed with the analysis of the fast mode only.
The solution for the eigen vector of the fast mode obtained upon substituting $\omega=kc$
into the equations for the perturbed electric field can be expressed now as,
\begin{eqnarray}
\delta E_1=\frac{k_2B_2D_{23}-k_2(k_2^2c^2+D_{22})[B_3+(\Omega_F/\omega)Rk_{||}B_p]}
{k_1(k_2^2c^2+D_{22})[B_3+(\Omega_F/\omega)Rk_{||}B_p]-k_2B_1D_{23}}\delta E_2,\label{dE1}\\
\delta E_3=-\frac{(k_1B_2-k_2B_1)(k_1k_2c^2+D_{21})}
{k_1(k_1k_2c^2+D_{21})[B_3+(\Omega_F/\omega)Rk_{||}B_p]-B_1D_{23}}\delta E_2.
\label{dE2}
\end{eqnarray}
Note that $D_{23}$ involves only first order terms.  To zeroth order we therefore obtain
$\delta E_1=-(k_2/k_1)\delta E_2$, implying that the perturbed charge and current 
density are indeed zero, as claimed by Punsly (2003).  However, there are corrections to the next
order.  By employing eqs. (\ref{pert3}), (\ref{dE1}) and (\ref{dE2}) one finds,
\begin{equation}
\delta \rho_e=-\frac{k_{||}j_p-k\rho_ec}{k_1c[B_3+(\Omega_F/\omega)Rk_{||}B_p]
- i4\pi(k_1B_1/k_{||}B_p)c[k_{||}j_p-k\rho_ec]}\delta E_2.
\label{rho-wav}
\end{equation}
Note that the second term in the denominator is smaller by a factor of $kR$ than the
first term, and can be neglected to the lowest order.
\subsection{Wave equations in the rotating frame}
To clarify the discrepancy between our results and those of Punsly (2003), we note that 
Punsly assumes the existence of a frame where the electric field vanishes and where 
Maxwell's equations preserve their form.  While a local Lorentz frame
where the electric field vanishes can be found (since $B^2-E^2>0$), one should keep in mind
that such a frame is only locally inertial, and there is no guarantee that the equations,
when written in terms of the local quantities should preserve their 
form.  To illustrate this point    
we now derive the wave equations in the rotating frame in the absence of a gravitational field.
For convenience we use spherical coordinates, and denote quantities in the rotating frame
by prime.  In the non-rotating frame the line element is, $ds^2=-dt^2+dr^2+
r^2(d\theta^2+\sin^2\theta d\phi^2$).  Performing a coordinate transformation to a frame 
rotating with angular velocity $\Omega_F$, viz., $d\phi^\prime=d\phi+\Omega_F dt$ 
(in units of c=1), 
one finds $ds^{\prime 2}=-dt^{\prime 2}+dr^{\prime 2}+r^{\prime 2}d\theta^{\prime 2}
+r^{\prime 2}\sin^2\theta^{\prime} (d\phi^{\prime}-\Omega_Fdt^{\prime })^2$.
All quantities can be transformed now to the rotating frame in the usual manner.  
It is readily seen that the only components of the electromagnetic tensor that 
are altered as a result of the transformation are those associated with the electric field:
$F^{\prime}_{t\theta}=F_{t\theta}-\Omega_FF_{\phi\theta}$, and 
$F^{\prime}_{tr}=F_{tr}-\Omega_FF_{\phi r}$.  Since for a force free flux tube 
$F_{t\theta}=\Omega_FF_{\phi\theta}$, and likewise for the radial component, we find that 
the electric field is indeed zero in the rotating frame. It is also easy to see that the 
only component of the 4-current that changes under the transformation is 
$j^{\prime \phi}=j^{\phi}+(r\sin\theta\Omega_F) j^t$, the other components remain unchanged.  In 
particular the charge density does not transform away.  Note that Gauss law is written in 
the rotating frame as,
\begin{equation}
-\frac{1}{r^2}\left[r^2 ( F^\prime_{t r}+\Omega_F F^\prime_{\phi r})\right]_{,r}
-\frac{1}{r^2\sin\theta}\left[\sin\theta( F^\prime_{t r}+ \Omega_F 
F^\prime_{\phi \theta})\right]_{,\theta}= 4\pi j^{\prime t}.
\end{equation}
Thus, the unperturbed charge density is given by the same equation as in the non-rotating
frame, as expected from its transformation law.  Amper's law becomes.
\begin{equation}
(F^{\prime}_{tr}+\Omega_F F^{\prime}_{\phi r})_{,t}+
\frac{1}{r^2\sin\theta}\left(\sin\theta F^{\prime}_{r\theta}\right)_{,\theta}=4\pi j^{\prime r},
\end{equation}
and likewise for the $\theta$ component.  Linearizing these equations would yield the wave 
equations in the rotating frame where the electric field vanishes.   
Upon re-defining the fields
$\psi_{r(\theta)}= F^{\prime}_{tr(\theta)}+\Omega_F F^{\prime}_{\phi r(\theta)}$,
(which are of course just the components of the electric field in the non-rotating frame)
one obtains exactly the same wave equations as in the non-rotating frame.

\section{Maxwell equations in Boyer-Lindquist coordinates}
The Kerr spacetime is a stationary, axisymmetric solution to the Einstein equations. 
In Boyer-Lindquist coordinates, $(t,r,\theta,\phi)$, the line element can 
be written in the form,

\begin{equation}
ds^2= -\alpha^2dt^2 + g_{\phi\phi}(d\phi+\beta dt)^2 + g_{rr}dr^2 + 
g_{\theta\theta}d\theta^2,
\end{equation}
where 

\begin{equation}
\alpha=\frac{\rho}{\Sigma}\sqrt{\Delta};\ \  \beta=-\frac{2aMr}{\Sigma^2};\ \ 
 g_{rr}=\frac{\rho^2}{\Delta};\ \  g_{\theta\theta}=\rho^2;\ \ g_{\phi\phi}\equiv\tilde{\omega}^2=
\frac{\Sigma^2}{\rho^2}\sin^2\theta
\end{equation}
with $\Delta=r^2+a^2-2Mr$; $\rho^2=r^2+a^2\cos^2\theta$; and 
$\Sigma^2=(r^2+a^2)^2-a^2\Delta\sin^2\theta$.  The parameter $a=J/M$ represents the specific 
angular momentum.  The units used above are geometrical units, in which $r$, $M$ and $a$ 
have units of length.

The electromagnetic field in curved spacetime is describe by the tensor $F^{\mu\nu}$
as usual.  It obeys Maxwell's equations,

\begin{eqnarray}
F^{\beta\alpha}_{;\alpha}=\frac{1}{\sqrt{-g}}(\sqrt{-g}F^{\beta\alpha})_{,\alpha}
=4\pi j^{\beta},\label{F=j}\\
F_{\alpha\beta,\gamma}+F_{\beta\gamma,\alpha}+F_{\gamma\alpha,\beta}=0,\label{F=0}
\end{eqnarray}
where $j^{\mu}$ is the generalized 4-current density.  The force-free condition is 
$F_{\mu\nu}j^\nu=0$.
In Kerr geometry Maxwell's equation
reduce to the rather simple form: 
 
\begin{equation}
F_{tr,\theta}+F_{r\theta,t}+F_{\theta t,r}=F_{\phi t,\theta}+F_{\theta\phi,t}
=F_{\phi t,r}+F_{r\phi,t}=0,
\label{Max-Homg}
\end{equation}
and 

\begin{eqnarray}
-\left[\frac{\Delta \sin\theta}{\alpha^2}(F_{tr}-\beta F_{\phi r})\right]_{,r}
-\left[\frac{\sin\theta}{\alpha^2}(F_{t\theta}-\beta F_{\phi \theta})\right]_{,\theta}=
4\pi\sqrt{-g}j^{t},\label{Max t}\\
\left[\frac{\rho^2}{\Delta\sin\theta}F_{t\phi}\right]_{,t}+
\left[\frac{\Delta \sin\theta}{\alpha^2}(\beta F_{tr}-\frac{g_{tt}}{g_{\phi\phi}} 
F_{\phi r})\right]_{,r}
+\left[\frac{\sin\theta}{\alpha^2}
(\beta F_{t\theta}-\frac{g_{tt}}{g_{\phi\phi}}F_{\phi \theta})\right]_{,\theta}=
4\pi\sqrt{-g}j^{\phi},\label{Max phi}\\
\left[\frac{\Delta \sin\theta}{\alpha^2}(F_{tr}-\beta F_{\phi r})\right]_{,t}+
\left(\frac{\Delta\sin\theta}{\rho^2} F_{r\theta}\right)_{,\theta}=4\pi\sqrt{-g}j^r, \label{Max r}\\
\left[\frac{\sin\theta}{\alpha^2}(F_{t\theta}-\beta F_{\phi \theta})\right]_{,t}-
\left(\frac{\Delta\sin\theta}{\rho^2}F_{r\theta}\right)_{,r}=4\pi\sqrt{-g}j^{\theta}.
\label{Max theta}
\end{eqnarray}

It is convenient to define the quantities $\psi_\theta=-E_\theta/\sqrt{\Delta}=
\Sigma(F_{t\theta}-\beta F_{\phi\theta})/(\rho^2\Delta)$;
$\psi_r=-E_r=\Sigma(F_{t r}-\beta F_{\phi r})/\rho^2$; 
$B_T=(\Delta\sin\theta/\rho^2)F_{r\theta}$; and 
$\tilde{j}^\phi=j^\phi+\beta j^t$, where $E_r$ and $E_\theta$ are the electric field 
measured by a ZAMO.  The independent variables are then: $\psi_\theta$, $\psi_r$, $B_T$,
and the toroidal component of the vector potential, $A_\phi$, which in steady state defines
magnetic flux surfaces.  In terms of these quantities the inhomogeneous Maxwell 
equations reduce to the form 
\begin{eqnarray}
\left(\Sigma\psi_r\right)_{,t}+\frac{1}{\sin\theta}B_{T,\theta}=4\pi\rho^2 j^r,\label{M1}\\
\left(\Sigma\psi_{\theta}\right)_{,t}-\frac{1}{\sin\theta}B_{T,r}=4\pi\rho^2 j^\theta,\label{M2}\\
\left(\frac{\rho^2}{\Delta \sin^2\theta} A_{\phi,t}\right)_{,t}-
\left(\frac{\Delta}{g_{\phi\phi}} A_{\phi,r}\right)_{,r}-
\frac{1}{\sin\theta}\left(\frac{\sin\theta}{g_{\phi\phi}} A_{\phi,\theta}\right)_{,\theta}=
4\pi\rho^2\tilde{j}^\phi.\label{M3}
\end{eqnarray}
The homogeneous Maxwell equation becomes
\begin{equation}
\left(\frac{\rho^2}{\Delta \sin\theta}B_T\right)_{,t}=\left(\frac{\rho^2\Delta}
{\Sigma}\psi_\theta\right)_{,r}-\left(\frac{\rho^2}{\Sigma}\psi_{r}\right)_{,\theta}
+\beta_{,r} A_{\phi,\theta}-\beta_{,\theta} A_{\phi,r},
\label{M5}
\end{equation}
and the force free condition is written as:
\begin{eqnarray}
\psi_r j^r+\Delta\psi_\theta j^\theta +A_{\phi,t}\tilde{j}^\phi=0,\label{FF1}\\
A_{\phi,r} j^r+A_{\phi,\theta} j^\theta +A_{\phi,t}j^t=0,\label{FF2}\\
\psi_r j^t-A_{\phi,r}\tilde{j}^\phi- \left(\frac{\rho^2}{\Delta \sin\theta}B_T\right)
j^\theta =0,\label{FF3}\\
\Delta\psi_\theta j^t-A_{\phi,\theta}\tilde{j}^\phi+ \left(\frac{\rho^2}
{\Delta \sin\theta}B_T\right)j^r =0.\label{FF4}
\end{eqnarray}


\begin{thebibliography}{99}
\bibitem[Blandford \& Znajek(1977)]{bz77} Blandford, R.D., \& Znajek, W.L., 1977,
  MNRAS, 179, 377 
\bibitem[Blandford]{b01} Blandford, R.D. 2001, Prog. Theor. Phys. Suppl., 147, 182
\bibitem[Blandford]{bz02} Blandford, R.D. 2002, in Lighthouses of the Universe, ed. M.
Gilfanov, R.A. Siunaiaev, \& E. Churazov (New York:Springer)
\bibitem[Bes]{bes} Beskin, V.S. 1997, Phys. Uspekhi, 40, 659
\bibitem[BK]{bk} Beskin, V.S. \& Kuznetsova, I.V. 2000, Nuovo Cimento B, 115, 795
\bibitem[]{Getal} Goodwin, P.S., Mestel, J., Mestel, L., Wright G. 2003, MNRAS, in press
\bibitem[k]{ks} Koide, S., Shibata, K., Kudoh, T., \& Meier, D. 2002, Science, 295, 1688
\bibitem[k]{k} Koide, S., 2003, Phys. Rev. D., 67, 104010
\bibitem[kom]{kom} Komissarov, S.S., 2001, MNRAS, 326, L41
\bibitem[Mest]{Mest} Mestel,, L. 1999, Stellar Magnetism. Clarendon Press, Oxford.
\bibitem[phy]{Phy} Phinney, E.S., 1983, Ph.D dissertation, University of Cambridge
\bibitem[Punsly]{P03} Punsly, B. 2003, ApJ, 583, 842
\bibitem[Punsly \& Coroniti(1990)]{PC90} Punsly, B., \& Coroniti, F.V. 1990, ApJ, 350, 518
\bibitem[Takahashi et al.(1990)]{takh90} Takahashi, M., Nita, S., Tatematsu, Y., \& Tominatsu, A.
        1990, ApJ, 363, 206 
\bibitem[Thorne, et al.(1986)]{tho86} Thorne, K.S., Price, R.H., \& Macdonald, 
D.A. 1986, Black Holes: The Membrane Paradigm. Yale University Press.
\bibitem[U]{} Uchida, T. 1997, MNRAS, 291, 125
\bibitem[van Putten(2001)]{mvp01a} van Putten, M.H.P.M., 2001, Phys. Rep., 345, 1
\bibitem[van Putten(2002)]{mvpl} van Putten, M.H.P.M. \& Levinson, A. 2003, ApJ, 584, 93
\end{thebibliography}
\end{document}